# Light localization in hollow core fibers with a complicated shape of the core – cladding boundary

*A. D. Pryamikov, A.S .Biriukov, G.K. Alagashev*

*Fiber Optics Research Center of Russian Academy of Sciences, Russia*

*e –mail:pryamikov@fo.gpi.ru*

*In this paper we present a theoretical and numerical analysis of light localization in hollow core microstructured fibers (HCMFs) with a complicated shape of the core – cladding boundary. The analysis is based on well - established models (for example, the ARROW (anti-resonant reflecting optical waveguide) model) and also on the models proposed for the first time. In particular, we consider local and nonlocal mechanisms of light localization in the waveguide structures with a determined type of discrete rotational symmetry of the core – cladding boundary. We interpret and analyze mechanisms of light localization in negative curvature hollow core microstructured fibers (NC HCMFs) and simplified HCMFs with a polygonal shape of the core – cladding boundary.*

## 1. Introduction

By now, a large variety of waveguide microstructures from holey fibers [1] to hollow core fibers with complicated microstructured claddings [2 - 4] have been developed. The corresponding mechanisms of light localization from a modified total internal reflection to a photonic band gap mechanism [1] have been proposed. The general approach to explaining the light localization mechanism in all solid and hollow core microstructured photonic band gap fibers with an 1D periodicity of the photonic crystal cladding is well known and it is called the ARROW mechanism [5]. The ARROW mechanism implies that the 1D periodicity in the photonic crystal cladding does not always give rise to the Bragg constructive interference necessary for the light localization in the core. If the wavelength of the localized radiation is short enough in comparison with the unit cell of the photonic crystal cladding $\lambda < \Lambda$ each high index layer can be considered as an individual scatterer. As a result, it is its optical properties and not the properties of the photonic crystal cladding as a whole that determine the transmission of the waveguide. Each of these layers can be considered as a planar Fabry – Perro resonator with its own resonance system while the impact of the cylindrical surface can also be taken into account [6]. In this case, the transmission band edges are determined by the transverse resonances of these individual cladding layers but not the Bragg resonances of the cladding. Inside the bands, light reflects effectively from each of the layers in an antiresonant regime and interferes in the air core while forming low leaky modes. According to the modern understanding, the ARROW mechanism can be used to explain the waveguide mechanism in all solid band gap fibers with a 2D photonic crystal cladding [6, 7]. Since the refractive index periodicity in the cladding is in the transverse direction, the resonances in the high index layers are described by the transverse component of the wavevector (linear momentum of the wave).

Recent studies have shown that two following factors play the key role in the process of light localization in HCMFs with a complicated cladding structure: the structure of a photonic crystal cladding [3, 4] (its antiresonant properties) and resonant properties of the hollow core which are determined by its size and the shape of the core – cladding boundary. In the case of HCMFs with the photonic crystal cladding [3] and a core – cladding boundary with continuous rotational symmetry, the light localization in the hollow core occurs as a result of the spectral overlapping of the resonances of the hollow core and antiresonant spectral range of the cladding [1] (photonic band gaps). In the case of HCMFs with a Kagome lattice photonic crystal cladding, the light localization mechanism is more complicated and not based on the photonic band gaps. At present, a generally accepted explanation of the waveguide mechanism in HCMFs with a Kagome lattice photonic crystal cladding is based on the inhibited coupling model [8, 9]. According to this model, the light is localized in the hollow core due to the low density of optical states in the Kagome lattice photonic crystal cladding, to a small value of the overlap integral between the air core modes and the cladding states, and due to a phase mismatch between them [4]. Also, the authors of [10] demonstrated that optical properties of HCMFs with a Kagome lattice photonic crystal cladding with a bigger pitch value than the wavelength are completely determined by geometrical parameters of the hollow core. At that, the total loss of such HCMFs didn't substantially decrease when additional photonic crystal layers were incorporated into the cladding.

In [11] the HCMF core – cladding boundary was fabricated in the form of a hypocycloid curve, this modification leading to a significant total loss reduction in narrow spectral ranges. In [12] the same authors demonstrated a possibility of drastic decrease in the total loss in a HCMF with Kagome lattice photonic crystal cladding by means of the hypocycloid core – cladding boundary with different curvature values. It is worth noting that the HCMF cladding structure, much like in the case of [10], didn't have any major impact on the loss level and became apparent only in small oscillations inside the transmission bands [12]. In this case, the strong localization of the air core modes occurred only due to their interaction with the hypocycloid core – cladding boundary.

It is clear from the above considerations that the structure of the microstructured cladding [3, 4] (in particular, its antiresonant properties) and the resonant properties of the hollow core determined by its size and shape play an important role in the process of light localization in HCMFs.

Thereupon, if the shape of the core – cladding boundary has a pronounced effect on optical properties of the hollow core, one can investigate a possibility of fabricating an HCMF without a complicated cladding structure and with such a shape of the core – cladding boundary which allows for a low loss light transmission. This problem was largely investigated in [13, 14]. In these works the waveguide structures didn't have a complicated cladding structure. In particular, in [13] the microstructure consisted of eight silica glass capillaries located symmetrically on the inside surface of silica glass mounting tube. In [14] the silica glass microstructure had a negative curvature (NC) core – cladding boundary in the form of joint "parachutes" ("ice – cream" in authors' terminology). Despite this simplification of the HCMF structure the transmission bands for the propagating radiation in the air core were observed experimentally up to wavelengths of 7 – 8 μm [15], where the silica glass was opaque. In [16] it was shown that optical properties of HCMFs with a negative curvature of the core – cladding boundary were determined to a great extent by the shape of the core – cladding boundary and, in particular, by the curvature of its elements. In the general case, the optical properties of NC HCMFs are determined by discrete rotational symmetry of the core – cladding boundary [17], the curvature of a single cladding element [12, 16] and its antiresonant properties [18, 19]. The calculations show that even in the case of the simplest NC HCMF structure [13] over 99.993% of radiation directed by the fiber can propagate in the air filling the air core [17]. It can then be concluded that the material properties of the cladding don't play such an important role as geometrical parameters of the microstructure construction.

In this work, we have considered idealized hollow core waveguide microstructures with a continuous or quasi continuous shape of the core - cladding boundary to analyze the effect of parameters of the core – cladding boundary on optical properties of HCMFs. Some of them have a polygonal shape of the core – cladding boundary with the azimuthal unit cell $\delta\varphi = 2\pi/N$, where $N$ is a number of the polygonal sides. In this case, the direction of local inner normal vector changes in a discrete way when passing from one azimuthal unit cell to another one. The other waveguide microstructures have a "the negative curvature" core – cladding boundary and the direction of local inner normal vector changes in a continuous way along the boundary of the unit cell with the direction changing discontinuously under transition to the neighboring elementary cell. As a rule, in this case the elementary cell is a semicircle with convexity directed to the microstructure center.

The analysis of such simplified and idealized waveguide systems provides insights into the mechanism of light localization in HCMFs with periodical boundaries in the azimuthal direction. As it will be shown below, the main difference between the waveguide mechanisms in HCMFs with a periodical shape of the core - cladding boundary in the azimuthal direction and in HCMFs with a circular shape of the core – cladding boundary comes down to a complication of the air core modes structure in different regions of the core – cladding boundary. As a result, it leads to a complication of the boundary conditions for the regions. In this case, to describe the waveguide mechanisms occurred in the complicated microstructures it is necessary to introduce the local boundary conditions and relevant local inner normal much like in the theory of optical microcavities with a complicated boundary shape [20, 21].

The rest of the paper is divided into 3 parts: Section 2, where we consider the difference between the waveguide mechanism in HCMFs with a determined type of the rotational symmetry of the core – cladding boundary and the ARROW mechanism; Section 3, where we consider the mechanism of the air core modes formation in NC HCMFs based on the local boundary conditions; and Section 4 where we give the conclusions.

## 2. Mechanism of light localization in waveguides with a polygonal shape of the core – cladding boundary

We start by comparing the mechanisms of light localization in HCMFs with a polygonal shape of the core – cladding boundary to those in HCMFs with a circular core – cladding boundary. Let us consecutively consider hollow core waveguide microstructures with circular and polygonal forms of the cross – section of the core – cladding boundary, where the number of the polygonal sides changes in the range of $N = 3 - 8$. Waveguide properties of the polygonal microstructures were first studied in [22, 23]. In these works the dispersion properties and loss spectra of the microstructures were investigated. In particular, it was shown that the polygonal shape of the core – cladding boundary led to asymmetrical Fano resonances with dispersion characteristics of the air core modes inside the transmission bands. The resonant coupling between the air core modes and core – cladding modes led to the resonant increase in the loss level of the transmission spectrum.

In contrast to [22, 23] we use the above comparison to understand the basic common factors of the air core mode field transformation in the microstructure walls as well as in different spatial domains of the waveguide microstructures depending on the rotational symmetry of the core – cladding boundary.

It is assumed that all microstructures are made of silica glass and have the same wall thickness. The air cavities diameters were selected in such a way that the effective mode areas are equal to each other in all cases. In Fig. 1 the leaky loss dependencies on wavelength for the fundamentals air core modes are shown in one transmission band with a center at a wavelength of 1.6 μm. The calculations were carried out by the finite element method (commercial packet Femlab 3.1) already used by us in [13]. Fig. 2 shows that the square microstructure has a minimal leaky loss, the pentagonal microstructure has a higher leaky loss comparable to that of the triangular microstructure. The octagonal microstructure has the highest leaky loss.

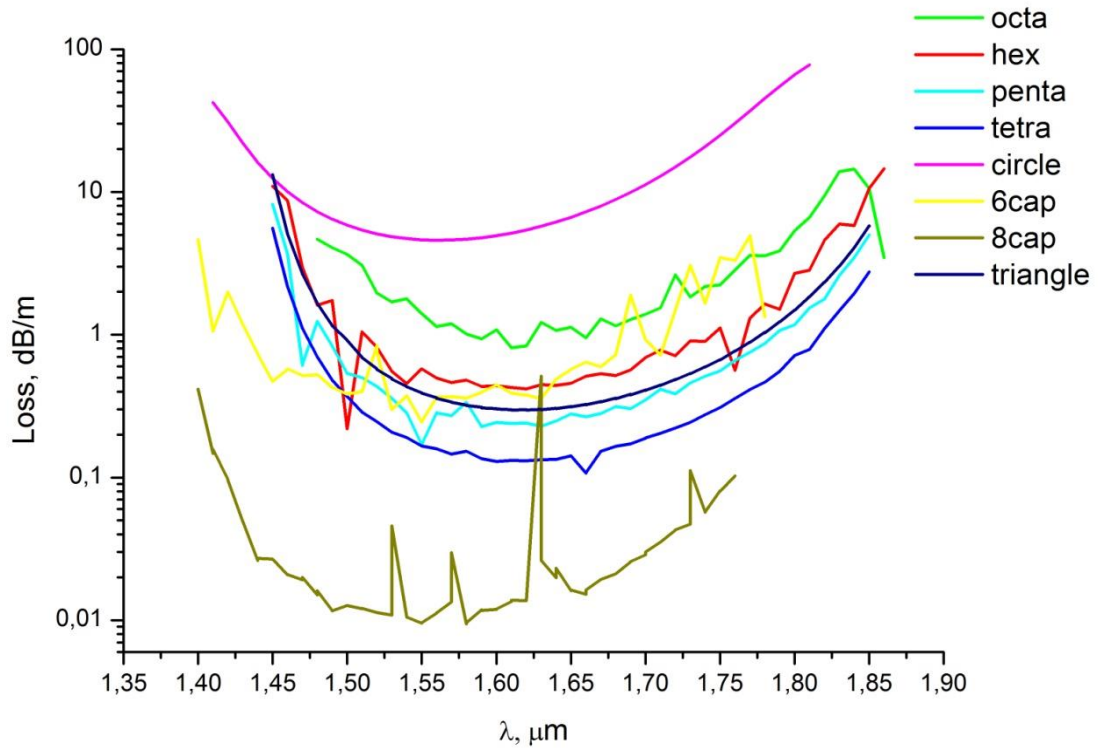

Fig. 1. Leaky loss for the waveguide microstructures with different shapes of the core – outer medium boundary

In addition, the fundamental mode of the circle microstructure has a maximal leaky loss (Fig. 1). The authors of [24] have come to the same conclusion. Moreover, it can be noted that all curves in Fig. 1 have a resonant behavior inside the transmission band with the exception of those corresponding to the circular and triangular microstructures. To explain the behavior of the spectral - loss curves in Fig. 1 let us consequently consider physical mechanisms leading to the light localization in the hollow cores of all the seven waveguide microstructures.

Localization of the air core modes in the dielectric tube is generally attributed to the ARROW mechanism [5]. The distribution of the absolute value of the electric field transverse component common to the ARROW mechanism is shown in Fig. 2 (right). The resonant condition for the fields in the tube wall is described as in the case of plane parallel Fabry – Perot system by the formula $k_t d=\pi m$, where $k_t$ is a transverse component of the wavevector, $d$ is the thickness of the tube wall and $m$ is an integer. The tube wall is a good reflector in the spectral ranges where this equality is not satisfied and the antiresonant regime of radiation propagation occurs in the hollow core.

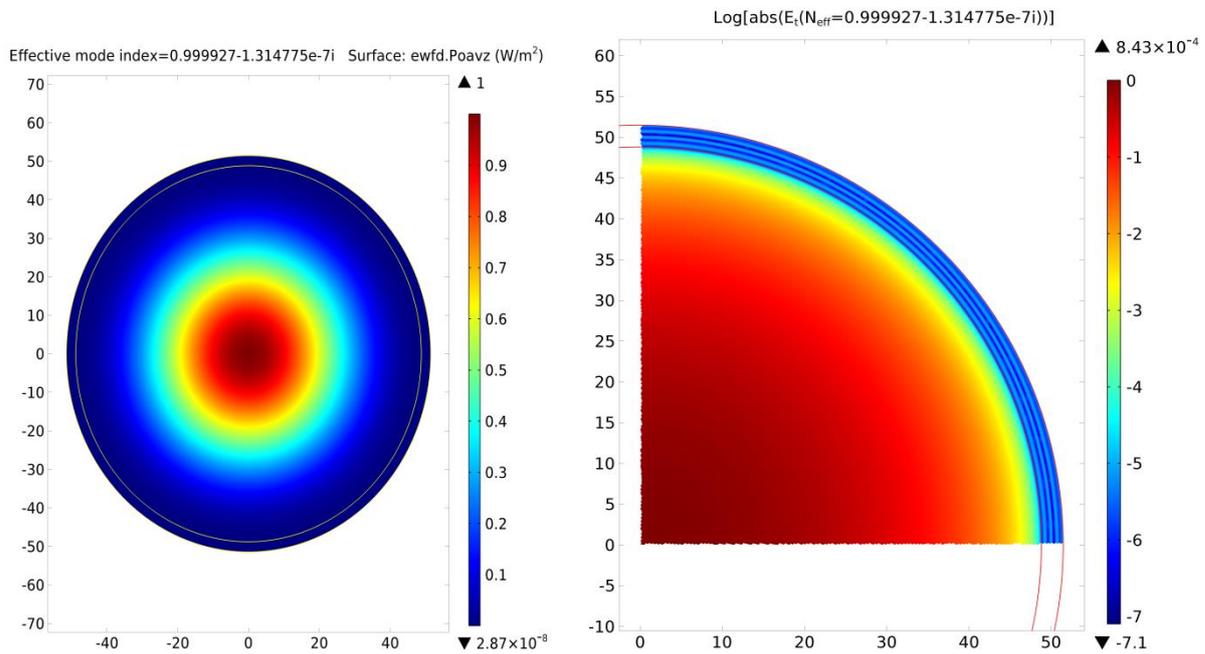

Fig. 2. The fundamental air core mode of the dielectric tube with a circular cross section (left) and a distribution of the absolute value of its electric field transverse components (right). As in the figures below, logarithm of absolute value of the electric field transverse component is shown on the right axis.

The $E$ and $H$ fields of the air core modes are obtained by solving the Helmholtz equation in each spatial region with the refractive index $n_i$:

$$\left(\vec{\nabla}^2 + \varepsilon_i k_0^2\right)\begin{bmatrix}\vec{E}\\\vec{H}\end{bmatrix} = 0, \qquad (1)$$

where $k_0=\omega/c=2\pi/\lambda$ is a wavenumber in free space, $\omega$, $\lambda$ are a cyclic frequency and a wavelength, and $\varepsilon_i = n_i^2$. In (1) the permeability µ is assumed to be equal to 1 in the optical spectral range.

$$\left(\Delta_\perp + k^2\right)V = 0 \qquad (2)$$

For the circular dielectric tube (Fig. 2) the equation (1) is analyzed in a cylindrical coordinate system ($r$, $\varphi$, $z$). The dependencies of all field components on time $t$ and the axial coordinate $z$ are the same and are written as $e^{i(\beta t - \omega t)}$. From (1), the equation for the axial components of the electromagnetic field can be obtained:

$V$ means $E_z$ or $H_z$; $k$ is a transverse wavevector $k = \sqrt{k_0^2 n_i^2 - \beta^2}$ in the medium with the refractive index $n_i$, $\beta$ is a propagation constant and $\Delta_\perp = \Delta - \partial^2/\partial^2 z$ is a transverse laplacian which is given in the cylindrical coordinates by:

$$\Delta_\perp = \frac{1}{r}\frac{\partial}{\partial r}r\frac{\partial}{\partial r} + \frac{1}{r^2}\frac{\partial^2}{\partial \varphi^2}.$$

The solutions to the wave equation (2) are written as cylindrical harmonics expansions:

$$E_z = \sum_{m=-\infty}^{+\infty} A_m F_m(r) e^{im\phi} e^{i\beta_m z},$$

$$H_z = \sum_{m=-\infty}^{+\infty} B_m F_m(r) e^{im\phi} e^{i\beta_m z}, \qquad (3)$$

where $A_m$, $B_m$ are the expansion coefficients and $F_m(r)$ are cylindrical functions which are determined by the physical conditions of a problem and by singularities of a considered spatial region. In (3) each term with number $m$ is a normal mode of the waveguide shown in Fig. 2 with its own propagation constant $\beta_m$ satisfying the corresponding dispersion equation. These dispersion equations are obtained from boundary conditions for tangential components of the air core modes fields. The common multiplier $e^{i\omega t}$ will be omitted for simplicity.

Using the above axial components, the transverse components of the air core modes fields can be given as [25]:

$$\vec{E}_t = \frac{i}{k^2}(k_0 \vec{z} \times \vec{\nabla}_t H_z + \beta \vec{\nabla}_t E_z),$$

$$\vec{H}_t = \frac{i}{k^2}(n_0^2 k_0 \vec{z} \times \vec{\nabla}_t E_z - \beta \vec{\nabla}_t H_z).$$

Correspondingly, the number of resonances in the tube wall at a given wavelength is described by the number of zeros of the $F_m(r)$ function. The energy of the air core mode leaks in the transversal direction and concentrates in one spatial channel as in the air core as well as in the tube wall. By a "spatial channel" we mean a geometrical configuration of the air core mode with a given azimuthal number $m$. The application of boundary conditions to the tangential components of the air core mode fields is univalent at the core – outer medium boundaries. The boundary conditions are the same along the core - outer medium boundary in the dielectric tube with a continuous rotational symmetry of the core – outer medium. The inner normal vector is directed along the radius at each point of the core – outer medium boundary. In this case, the boundary conditions for tangential components of electric and magnetic fields (1) are written as:

$$E_{zi} = E_{zj},\ E_{\varphi i} = E_{\varphi j},\qquad(4)$$

$$H_{zi} = H_{zj},\ H_{\varphi i} = H_{\varphi j},,$$

where $i$ and $j$ are indices corresponding to touching mediums in the waveguide structure. The mechanism of light localization in the hollow core of the dielectric tube (ARROW) is not local in this case and is applicable to the whole microstructure perimeter.

It will be shown below that for waveguide microstructures with a discrete rotational symmetry of the core – outer medium boundary the mechanism of the air core mode formation can be both nonlocal (as for dielectric tubes) and determined by the geometrical parameters of the individual boundary elements and the type of the boundary rotational symmetry. For microstructures with a discrete rotational symmetry of the core – cladding boundary the boundary conditions for tangential components of electromagnetic fields becomes much more complicated. This in turn leads to a complication of the fields' structure in the vicinity of the core boundary as well as to a complication of the one in the microstructure wall despite the wall thickness is equal to that of the dielectric tube. This dissimilarity between spatial structures of the electromagnetic fields at the core boundary and in the microstructure wall on the one hand, and those of the dielectric tube on the other, occurs due to a type of the rotational symmetry of the core – cladding boundary and, as a consequence, due to different types of the Helmholtz equation (1) in different spatial regions of the microstructure.

As in the case of the resonant cavity with a deformed boundary described by an equation $r = R(\varphi)$ [20, 21] the air core mode fields of the polygonal microstructures can be expanded into series of cylindrical harmonics [21]. The axial components of the air core mode fields must be governed by the Helmholtz equation (2) in the air core and in the outside medium and can be expressed in expanded forms:

$$E_z = \sum_{m=-\infty}^{+\infty} A_m F_m^{core}(r) e^{im\phi} e^{i\beta z}, \qquad r < R^{core}(\varphi)$$

$$E_z = \sum_{m=-\infty}^{+\infty} B_m F_m^{out}(r) e^{im\phi} e^{i\beta z}, \qquad r > R^{out}(\varphi). \qquad (5)$$

In (5) $R(\varphi)$ is a function of the azimuthal angle which describes the shape of inner $R^{core}(\varphi)$ and outer $R^{out}(\varphi)$ boundary of the microstructure wall in cylindrical coordinates originating at the microstructure center. The magnitudes $A_m$ nd $B_m$ are amplitudes of the axial components of the air core mode fields in the air core and in the outside medium, correspondingly. As in the case of (3), each term of sums (5) is a solution of the Helmholtz equation (2) in corresponding spatial regions.

The axial components of the magnetic fields in these spatial regions are expanded in the same way.

However, in contrast to the solution (3) no term of the sums (5) individually satisfies corresponding boundary conditions for tangential components of the fields. Thus, the fields of the individual air core modes in polygonal microstructures can only be given as a superposition of orthogonal modes. At that, the comparison of (5) and (2) shows that the introduction of the complicated shape of the core - cladding boundary described by the equation $r = R(\varphi)$ leads to a degeneracy of axial components of the fields.

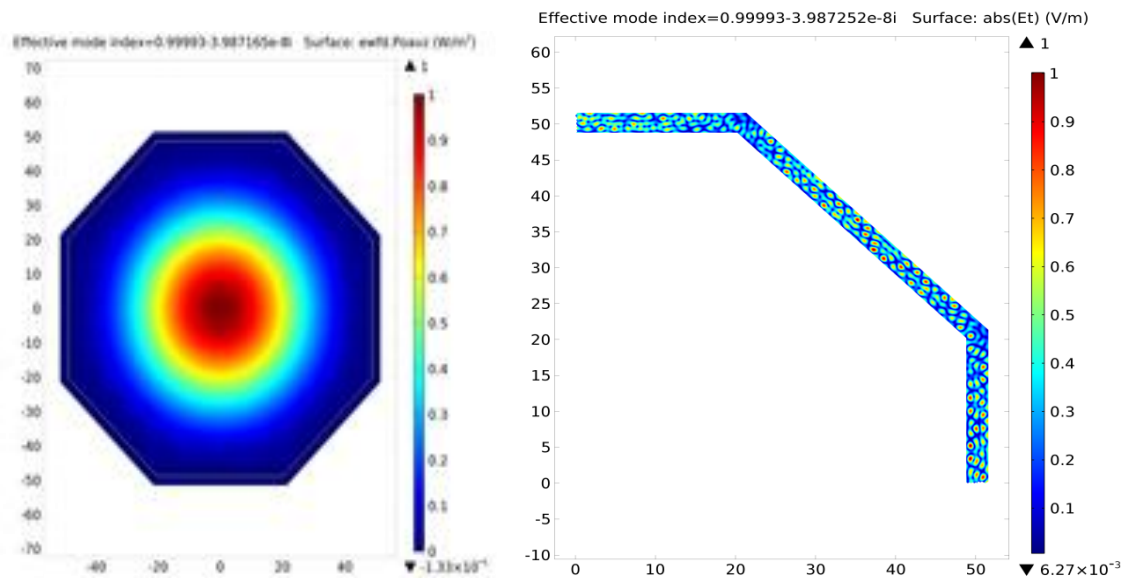

Fig. 3. The fundamental air core mode of the octagon microstructure (left) and a distribution of the absolute value of its transverse electric field components in the microstructure wall (right).

It occurs when a formally infinite number of azimuthal numbers *m* correspond to one propagation constant of the considered air core mode *β*. In view of the completeness of the orthogonal basis of the expansion terms (5) it is exact at $r < R^{core}(\varphi)$ or at $r > R^{core}(\varphi)$. According to Rayleigh's hypothesis [26] the expansions of the fields (5) can be continued analytically to the boundary region and to the microstructure wall. In work [27] it was shown that under high values of the boundary deformation (critical deformation) the hypothesis breaks down due to a non-analytic form of the expansion in the $R^{core}(\varphi) < r < R^{out}(\varphi)$ region. For the considered polygonal microstructures the significant dissimilarity of their boundaries from the circular shape leads to a modification of solution in the form (5) as a complete set of normal modes. Solutions of the Helmholtz equation in the microstructure wall region become much more complicated.

The equation (1) for the space region corresponding to a wall of the polygonal microstructure with the periodical azimuthal dependence $R(\varphi) = R(\varphi+2\pi p/N)$, where *N* is a number of sides and *p* is an integer, is written as:

$$\left(\vec{\nabla}^2 + \varepsilon_{wall}(R(\varphi))k_0^2\right)\begin{bmatrix}\vec{E}\\\vec{H}\end{bmatrix} = 0, \qquad (6)$$

where

$$\varepsilon_{wall}(\varphi + \frac{2\pi p}{N}) = \varepsilon_{wall}(\varphi).$$

Since the dependence of solutions (6) on the longitudinal coordinate *z* is the same as in (1) the axial components of electromagnetic fields of the air core modes must be governed by the scalar equation (2) with the substitution of constant dielectric permittivity for that with a periodic azimuthal dependence (6). In this case, the expansions for the axial field components become more complicated compared to (5) as it follows from the general conclusions of theory of photonic – crystal microstructures with discrete rotational symmetry [28]. The propagation constant *β* of the individual air core mode experiences additional degeneracy due to the discrete rotational symmetry of the microstructure wall with an infinite number of azimuthal numbers $m \pm Np$ now corresponding to the propagation constant [28]. All components of the air core modes fields are formed as a harmonics superposition with new azimuthal momentum components $m \pm Np$ [28]. As it was mentioned above, the equation for axial components of the fields retains an unperturbed form in the hollow core and in the outside medium with a dielectric permittivity $\varepsilon=1$ and its solutions are given by expressions (5).

Fig. 3 – 4 (right) show that a distribution of the absolute value of transverse components of the fundamental mode electric field in the microstructure wall is a result of complicated interference of cylindrical harmonics. This phenomenon has nothing to do with the ARROW distribution occurring in the dielectric tube wall (Fig. 2). The air core modes of the octagonal and pentagonal microstructures have an azimuthal periodical distribution of the absolute value of the electric field transverse components in the region of the inner wall boundary (Fig. 5) (where the distance from the boundary to the waveguide axis is minimal). This point to an interference between a degenerate set of harmonics with azimuthal numbers $m \pm pN$ and harmonics being part of the sum (5). Note that besides a stable, interference pattern , periodical in azimuthal direction, there are also random

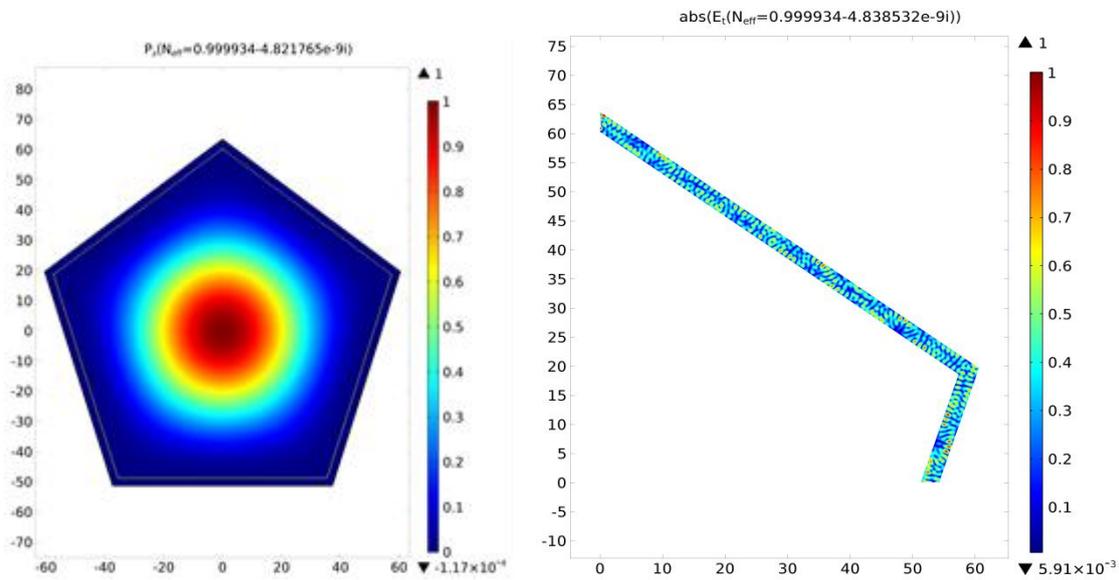

Fig. 4. The fundamental air core mode of the pentagonal microstructure (left) and a distribution of the absolute value of its transverse electric field components in the microstructure wall (right).

singularities of the electric field distribution in the angular domains of the core – cladding boundary (Fig. 5(right)).

Note that publications on the theory of light scattering in microcavities with a complicated shape of the core – cladding boundary [29] interpret the phase factors $e^{im\varphi}$ and $e^{-im\varphi}$ as traveling waves in the azimuthal direction: one wave goes clockwise while the other goes counterclockwise order (CW and CCW waves).

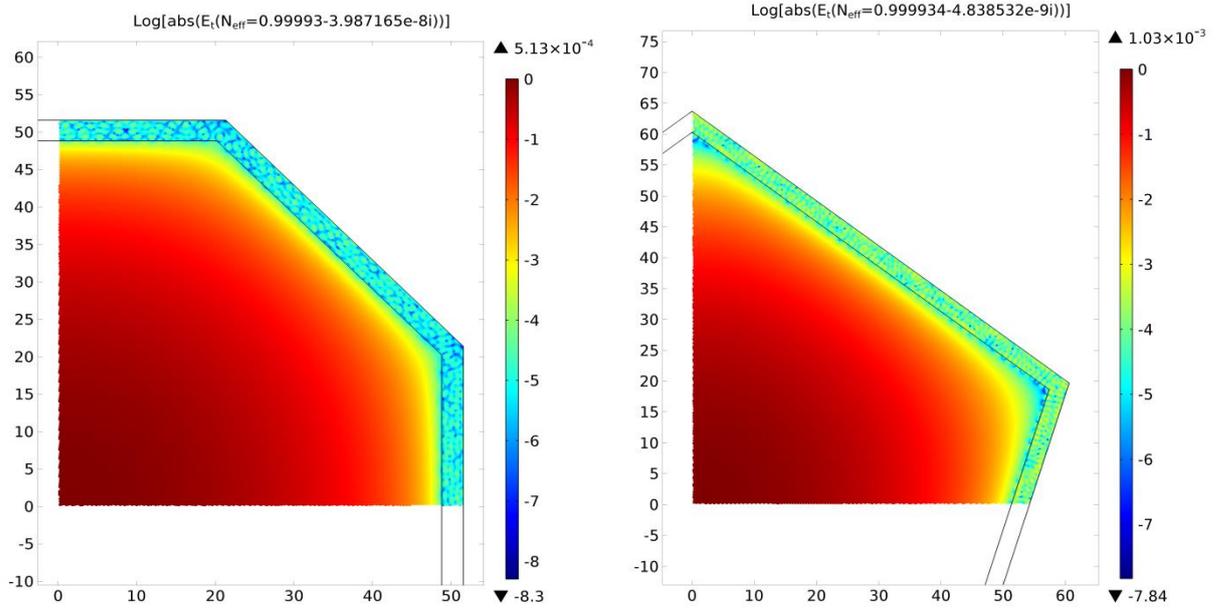

Fig. 5. The distribution of the absolute value of transverse components of the air core electric fields as in the vicinity of the core – cladding boundary and in the wall of octagonal (left) and pentagonal (right) microstructures (logarithmic scale).

This hypothesis was used in work [30] devoted to the Sagnac effect in a rotating microcavity and in [31] where a process of cyclic Sommerfeld waves formation was described. For polygonal waveguide microstructures, (Fig. 3, 4) given a finite thickness of the core – cladding boundary, the propagating CW and CCW azimuthal waves $e^{im\varphi}$ and $e^{-im\varphi}$ can be considered as an analog of propagating plane waves $e^{ikz}$ и $e^{-ikz}$ in an infinite one dimensional (1D) photonic crystal (PC) with a spatial period $\Lambda$. Then, the polygonal core – cladding boundary can be considered as an azimuthal photonic crystal with a spatial characteristic $2\pi p / N$, where $p$ is an integer [32]. In this case, such resonant effects as Bragg scattering should also occur in the azimuthal photonic crystal.

The behavior of the fields which are determined by (6) and interfere in the region of the core – cladding boundary can be analyzed with the coupling modes theory as a way to describe interference in an 1D photonic crystal [33]. Let us assume that $N \gg 1$. Then, the modulation of permittivity of the core – wall boundary and the microstructure wall $\varepsilon(R(\varphi))$ can be presented as a permittivity perturbation of the above circular dielectric tube. In this case, the permittivity of the core – cladding boundary can be written as:

$$\varepsilon = \varepsilon^{tube}(r) + \Delta\varepsilon(r,\varphi)$$

where $\varepsilon^{tube}(r)$ is the wall permittivity and $\Delta\varepsilon(r,\varphi)$ is its perturbation, while:

$$\Delta\varepsilon(r,\varphi) \neq 0 \text{ при } R^{core}(\varphi) \leq r \leq R^{out}(\varphi).$$

In general terms, the perturbation $\Delta\varepsilon(r,\varphi)$ can be expanded into a series:

$$\Delta\varepsilon(r,\varphi) = \sum_{p=-\infty}^{+\infty} \varepsilon_p(r) e^{iNp\varphi} . \qquad (7)$$

Based on the small value of the perturbation we can assume that the field amplitudes change slowly in the azimuthal direction thus enabling the use of the slow changing amplitude method [33]. The modes of expansion (5) form a complete orthonormal set of functions while the expansion is changed with the introduction of the discrete rotational symmetry of the core – cladding boundary (6). The perturbation inserted by the discrete rotational symmetry of the core – cladding boundary leads to coupling between the normal modes in the microstructure wall and to fields redistribution inside it (Fig. 5). In this case, taking into account (5), the general solution for the axial components of the air core modes fields can be written as:

$$E_z = \sum_q \sum_{m=-\infty}^{+\infty} A_m(\varphi) F_m^{wall}(r) e^{im\varphi} e^{i\beta_q z}$$

$$H_z = \sum_q \sum_{m=-\infty}^{+\infty} B_m(\varphi) F_m^{wall}(r) e^{im\varphi} e^{i\beta_q z}, \qquad (8)$$

where $q$ is the mode index. The expressions (8) show that the orthogonality condition for the normal modes (3), (5) forming the fields of the air core modes is modified in the microstructure wall because the amplitudes of the expansion coefficients are functions of the azimuthal angle which fails to satisfy the orthogonality condition. Substituting (8) in (6) we get an equation connecting the coefficients of different azimuthal orders:

$$\frac{1}{r^2} \sum_{m=-\infty}^{+\infty} \left( \frac{\partial^2 C_m(\varphi)}{\partial \varphi^2} + 2im \frac{\partial C_m(\varphi)}{\partial \varphi} \right) F_m^{wall}(r) e^{im\varphi} e^{i\beta_q z} + k_0^2 \sum_{l=-\infty}^{+\infty} \Delta\varepsilon C_l(\varphi) F_l^{wall}(r) e^{il\varphi} e^{i\beta_s z} = 0,$$

where by $C_m$ (and $C_l$) are the amplitudes of electric and magnetic components of the fields ($A_m(\varphi)$ or $B_m(\varphi)$). Further, applying the basic relation of the slow changing amplitude method to the amplitude azimuthal dependence:

$$\frac{\partial^2 C_m(\varphi)}{\partial \varphi^2} << 2im \frac{\partial C_m(\varphi)}{\partial \varphi},$$

and substituting the expansion (7) we obtain the equation:

$$\frac{2i}{r^2}\sum_{m=-\infty}^{+\infty}m\frac{\partial C_m(\varphi)}{\partial\varphi}F_m^{wall}(r)e^{i(m\varphi+\beta_q z)}=-k_0^2\sum_{l=-\infty}^{+\infty}\sum_{p=-\infty}^{+\infty}\varepsilon_p C_l(\varphi)F_l^{wall}(r)e^{i((l+Np)\varphi+\beta_s z)}\ . \qquad (9)$$

Then, multiplying (9) by $F^{wall*}(r)e^{-im\varphi}e^{-i\beta_q z}$ and performing integration over the cross – section of the microstructure wall we get:

$$\frac{\partial C_m(\varphi)}{\partial\varphi}=\frac{ik_0^2}{2mL}\sum_{l=-\infty}^{+\infty}\sum_{p=-\infty}^{+\infty}K_{lm}^p C_l(\varphi)e^{i(l+Np-m)\varphi}e^{i(\beta_s-\beta_q)z}\ , \qquad (10)$$

where $K_{lm}^p=\int_{R_{min}}^{R_{max}}F_l(r)\varepsilon_p(r)F_m^*(r)r^3 dr$ and $L=\int_{R_{min}}^{R_{max}}F_l^{wall}(r)F_m^{*wall}(r)rdr$, and $R_{min}$ and $R_{max}$ are the minimal and maximal values of $R^{core}(\varphi)$ and $R^{out}(\varphi)$, correspondingly.

As we can see from (10), the resonant coupling between the harmonics has two phase matching conditions: one is for the axial (linear) momentum component of the harmonics and the other is for azimuthal component of the harmonics. The conditions are commonly called "kinematic conditions". The second condition for the resonant coupling is called a "dynamical condition". It is determined by the inequality $K_{lm}^p\ne 0$ and depends on the polarization and spatial configuration of the interfering modes.

The equation (10) shows that the coupling occurs between the normal harmonics (5) whose interference is responsible for the air core mode formation when the periodic perturbation of the core – cladding boundary is inserted in the azimuthal direction. The coupling leads to a periodic mode field distribution along the core – cladding boundary (Fig. 5). The interference leads to the spatial redistribution of the mode energy flowing out of the air core and, consequently, to the decrease in the waveguide loss compared to the circular dielectric tube loss (Fig. 1).

Interfering harmonics have two types of resonant (kinematic) coupling which can occur at the core – cladding boundary and in the microstructure wall. The first type of coupling is determined by the kinematic condition for the azimuthal momentum components of the interacting harmonics $m=Np\pm l$ assuming that both harmonics belong to the same air core mode with the propagation constant $\beta_q$. The azimuthal number $Np$ in (7) leads to coupling of the resonant normal harmonics (5) with a complicated interference pattern occurring for the air core modes in the microstructure wall and at the core – cladding boundary (Fig. 5). Here, the number $Np$ can be considered as an analog of reciprocal lattice vector in 1D PCs [33]. In this case, for example, the Bragg type resonant coupling between the harmonics with the azimuthal dependencies $e^{im\varphi}$ and $e^{-im\varphi}$ can be realized under fulfilling the condition $m = Np/2$. As it was mentioned above, this fact leads to a phase mismatch between the fields in the hollow core and in the outside spatial region and, as consequence, to the configuration change of the space channels of the air core mode leakage. For the first time, the necessity to consider the azimuthal phase matching condition $m = Np \pm l$ to explain the additional loss origin in transmission bands of polygonal tube fibers was reported in [22].

The second resonant coupling may occur between an individual harmonic from the expansion (5) with the propagation constant $\beta^{air}/k_0 < 1$ and the microstructure wall mode, where the propagation constant is $\beta^{glass}/k_0 \approx 1$ when the condition $m = Np \pm l$ is fulfilled. In this case, the phase factor $e^{i(\beta_s - \beta_q)z}$ in (10) is close to a unit which leads to fulfilling of an additional phase matching condition for the propagation constants $\beta^{air} \approx \beta^{glass}$ (anticrossing). This resonant coupling between the two modes belonging to spatial regions with different refractive indices leads to a strong leakage of the air core modes to the outside region through the open space channel at a given wavelength. The transmission spectrum of the polygonal microstructures becomes "broken" (Fig. 1). A similar phenomenon is observed as in HC MFs with the Kagome lattice cladding [34] as well as for HC MFs with a curvilinear core – cladding boundary [17]. It was also observed that the number of resonances inside the transmission band (Fig. 1) considerably depends on the cross section shape of the core – cladding boundary, in particular, on its type of rotational symmetry. Lower number $N$ leads to a lower number of resonances inside the transmission band. We reported similar results in [17] for HC MF with a negative curvature core – cladding boundary when considering the process of the core - cladding boundary modes excitation with a determined type of rotational symmetry. The excitation of the curvilinear boundary modes with a specific type of discrete rotational symmetry also led to the resonant irregularity of the HC MF's transmission spectrum. For square waveguide microstructure the number of resonances inside the transmission band is considerably lower than for polygonal waveguide microstructures with a larger number of sides. The transmission band of the triangular microstructure is as smooth as that of circular dielectric tube (Fig. 1). It is demonstrated that the process of light localization in these cavities is mostly due to the ARROW mechanism which is related only to transverse resonances – antiresonances in the microstructure wall.

If the spectral loss dependencies for the polygonal waveguide microstructures with a small number of sides ($N < 5$) are determined by the ARROW mechanism rather than mechanism of azimuthal photonic crystal mechanism then the electromagnetic filed distributions in the microstructure walls should be different from the ones for the polygonal microstructures with a large number of sides ($N > 5$). Fig. 5 shows that the fields distributions in the microstructure walls with $N = 5$ и 8 is periodic and continuous along the whole perimeter of the boundary. Singularities in the behavior of the air core mode electric field occur only in the angular domains of the microstructures with $N = 5$. Corresponding changes in behavior of the fields in the microstructure wall with $N = 3, 4$ and, consequently, the change in the light localization mechanism are shown in Fig. 6 – 7(right). The particular characteristic of the microstructures with $N = 3, 4$ is boundary regions with markedly different field distributions (Fig. 6).

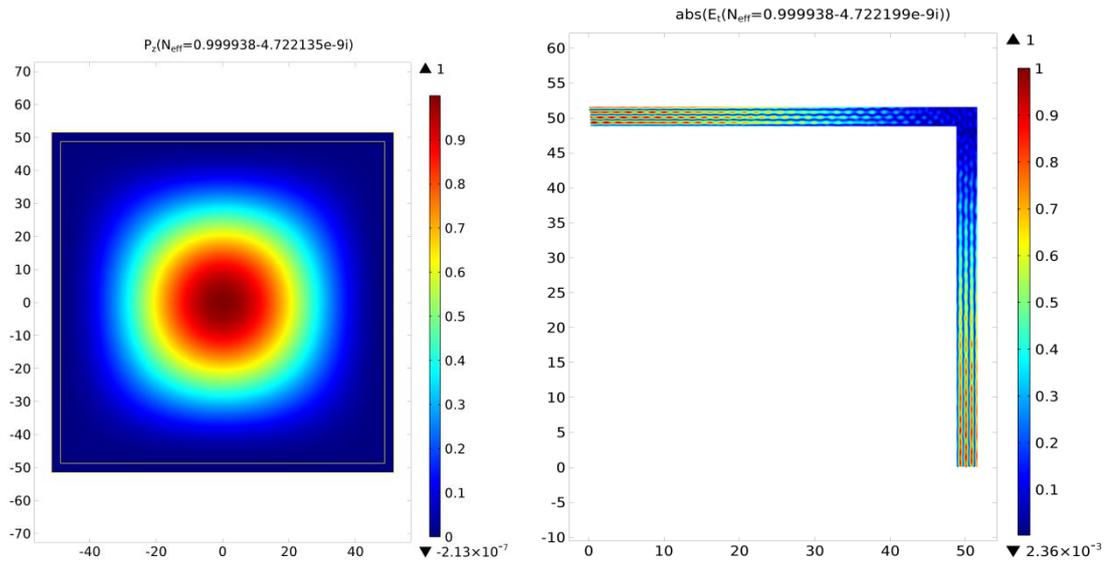

Fig. 6. The fundamental air core mode of the square microstructure (left) and a distribution of the absolute value of transverse components of the air core mode electric field in its wall (right).

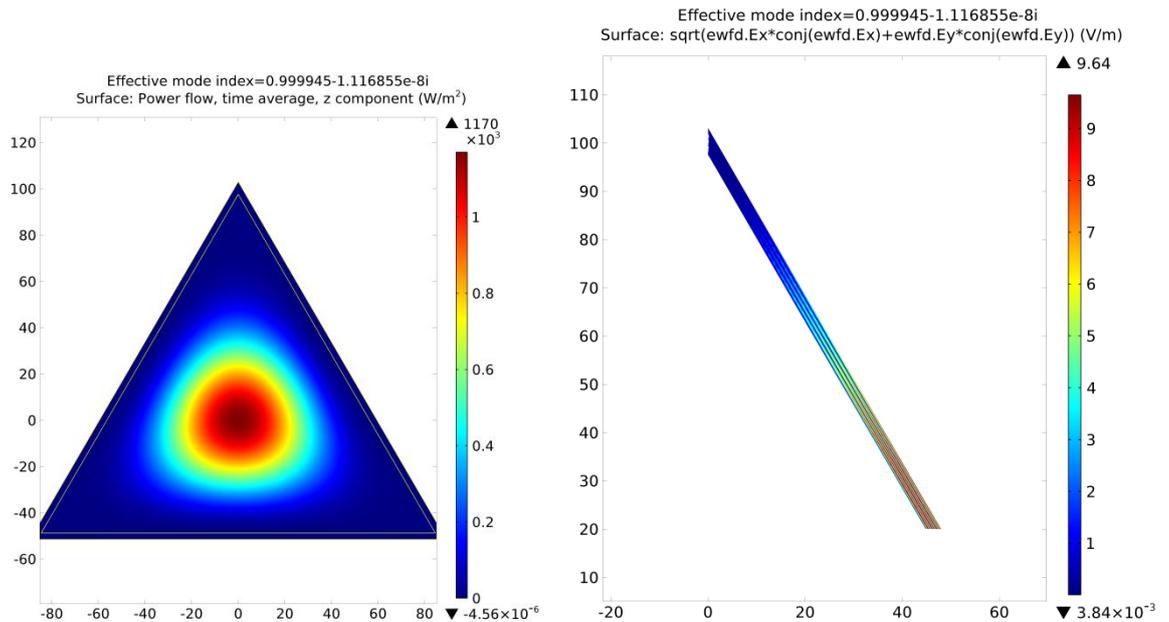

Fig. 7. The fundamental air core mode of the triangular microstructure (left) and a distribution of the absolute value of transverse components of the air core mode electric field in its wall (right).

The fields structure in the microstructure wall is no longer continuous and has regions with a field distribution typical for the ARROW mechanism. Such regions are located at a minimum distance from the square center.

Since the impact of the azimuthal photonic crystal mechanism on the formation and localization of the air core modes is significantly weakened in these regions of the core - cladding boundary it occurs only in the segments of the microstructure walls closest to the angular regions of the core – cladding boundary.

The singularities in the behavior of the field distributions in the angular regions of the microstructure walls with $N < 5$ manifest themselves much more distinctly compared to microstructures with $N > 4$. The boundary conditions become local and cannot be applied to the whole perimeter of the microstructure wall. They are determined by the values and signs of inner local normal vector components under displacement of the local normal vector along the inner boundary of the microstructure wall. For triangular microstructure (Fig. 7) the local ARROW mechanism effect is the most pronounced. Note that the square cavity has a lower waveguide loss compared to the triangular one, mostly due to a lower value of the boundary perimeter. Now, let us apply the results obtained above to explain the mechanism of light localization in waveguide microstructures with a negative curvature core – cladding boundary.

## 3. Mechanism of light localization in waveguide microstructures with a negative curvature core – cladding boundary

There is a strong resemblance between the mechanisms of mode formation in HC MFs with a negative curvature core – cladding boundary and HC MFs with a polygonal core – cladding boundary. In the former case, the mechanism of light localization in the cavity is related as to the interference of individual CW and CCW azimuthal propagating waves in the azimuthal photonic crystal mode as to the local ARROW mechanism. Let us consider two hollow core microstructures with a negative curvature core – cladding boundary but with a different discrete rotational symmetry of this boundary, for example, with number $N = 6$ and 8 capillaries in the cladding (Fig. 8, 9). The waveguide losses of the microstructures are shown in Fig. 1. The permittivity profile of the core – cladding boundary as in the case of polygonal microstructures is a function of the azimuthal angle. For the case under consideration the air core modes interact mostly with the core – cladding boundary in the regions closest to the microstructure line of symmetry. The air core modes of the compared microstructures have an azimuthal periodic structure analogous to that of polygonal microstructures (Fig. 5 and Fig. 10). Stated differently, the discrete rotational symmetry factor of the core – cladding boundary for HC MFs with a negative curvature core – cladding boundary also plays a crucial role in the interference of individual harmonics forming the air core modes. As for square and triangular cavities (Fig. 6 and 7) a sharp fall in electric field intensities in the angular domains of the core – cladding boundaries is observed (Fig. 10). Conspicuous is the fact that the distribution of the absolute values of transverse electric field components in the microstructure wall with $N = 8$ (Fig.8) is similar to those in microstructures shown in Fig. 6 and 7. Here, there is also the ARROW mechanism with a local radial field distribution in the boundary regions closest to the microstructure center. The waveguide loss of the microstructure air core modes is several orders of magnitude lower than the loss in the polygonal microstructures (Fig. 1). The distribution of absolute values of transverse electric field components in the microstructure wall with $N = 6$ (Fig.9) is similar to the ones shown in Fig. 3 and 4. Harmonics forming these fields interfere both in the azimuthal photonic crystal regime and in the local ARROW mechanism regime. The former mechanism is predominant. For this reason, the waveguide loss in this microstructure is much higher than in the microstructure with $N = 8$ and is comparable to the waveguide loss in the hexagonal microstructure (Fig. 1).

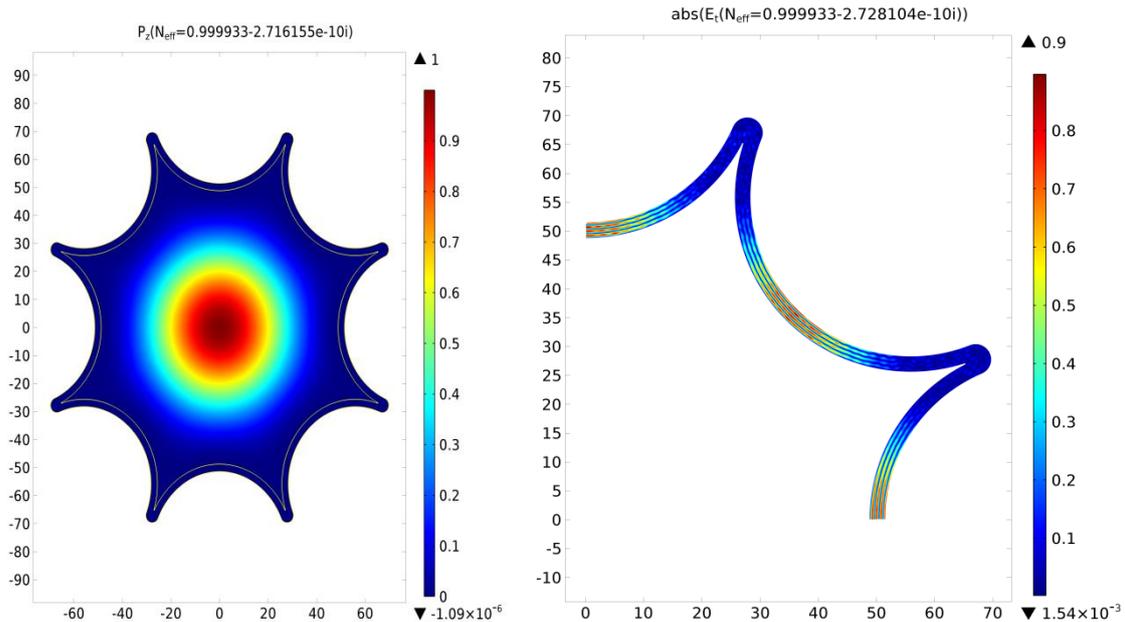

Fig. 8. The fundamental mode of the HC MF with a negative curvature core – cladding boundary ($N = 8$) (left) and a distribution of the absolute value of transverse electric field components in the microstructure wall (right).

Thus, both the light localization in HC MFs with a curvilinear core – cladding boundary and the light interaction with the boundary wall occurs due to the two waveguide mechanisms described in Section 2. As the balance between these two mechanisms of light localization shifts either the regime of strong light localization (the local ARROW) (Fig. 8) or the regime for polygonal microstructures (Fig. 9) can occur.

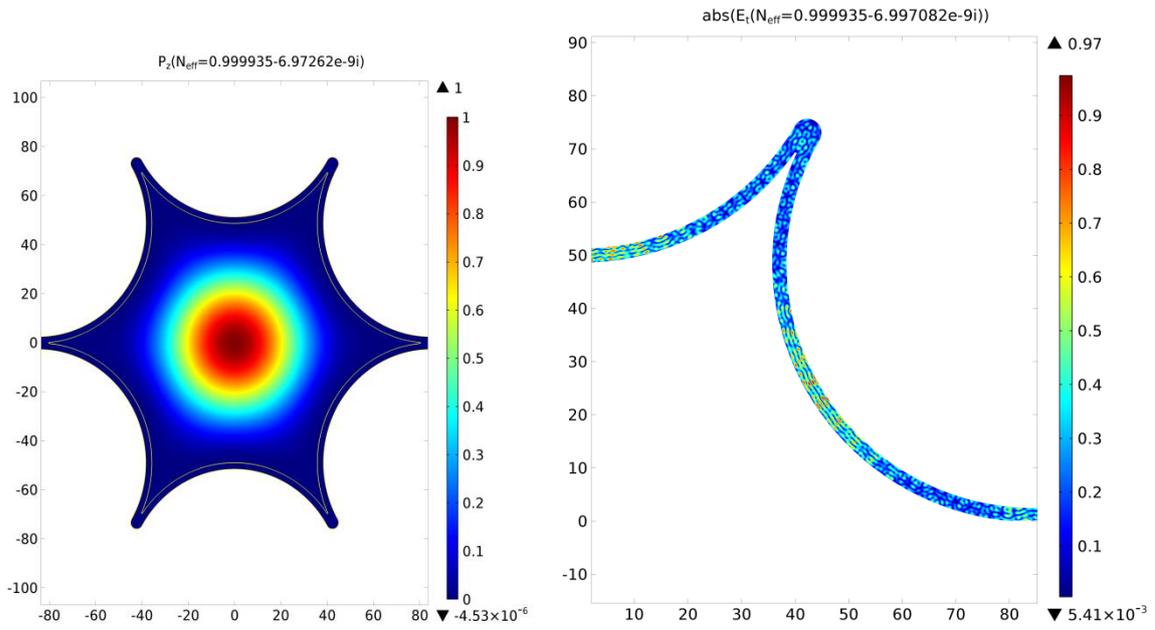

Fig. 9. The fundamental mode of the HC MF with a negative curvature core – cladding boundary ($N = 6$) (left) and a distribution of the absolute value of transverse electric field components in the microstructure wall (right).

In particular, this balance can be managed by changing the curvature of an individual cladding element which results in displacement of the air core mode fields from the spatial domains between the boundary elements.

Let us consider in more detail the mechanism of the field displacement from the angular domains of the waveguide microstructures (Fig. 10) based on local boundary conditions that used in theory of optical microcavities with a complicated shape of the boundary [20, 21]. Using the Helmholtz equation (6) for the axial field components we write them in cylindrical coordinates. At that, the coordinates origin coincides with the axis of the microstructure symmetry and we take the microstructure boundary to be continuous or quasi continuous. Let us determine a vector tangent to the inner surface of the core – cladding boundary by a vector product $\vec{t} = \vec{z} \times \vec{n}$, where unit vector $\vec{z}$ is directed in the axial direction and $\vec{n}(r(\varphi))$ is a unit local normal vector to the surface. Note that any vector in the core – cladding boundary region can be expressed via the vector tripod $(\vec{z}, \vec{t}, \vec{n})$. Then, the derivatives with respect to transverse coordinates in the expressions of the transverse field components in Section 2 are changed by directional derivatives:

$$\frac{\partial}{\partial n} = (\vec{n}\vec{\nabla}_t), \ \frac{\partial}{\partial t} = (\vec{t}\vec{\nabla}_t).$$

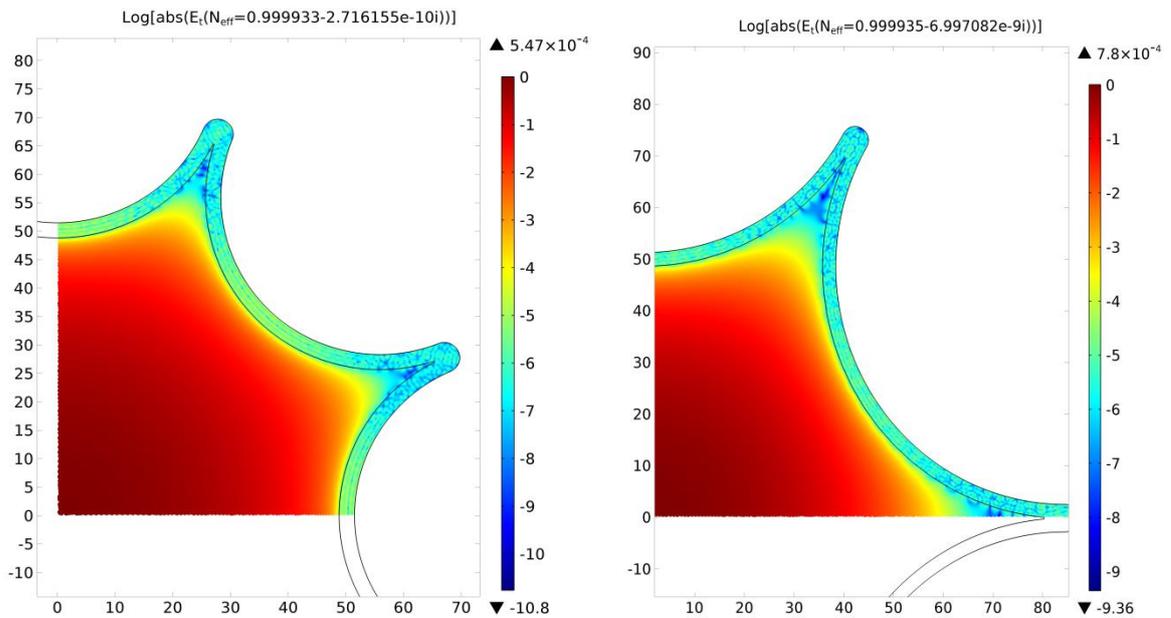

Fig. 10. The distribution of the absolute value of transverse components of the air core electric fields for the waveguide microstructures with $N = 8$ (left) and $N = 6$(right) (logarithmic scale).

The transverse component of the wavevector becomes local and takes the form of $\gamma^2 = \varepsilon(r(\varphi))k_0^2 - \beta^2$.

Typically, the boundary conditions require the tangent electromagnetic field components to be continuous on both sides at each point of the interface. For nonlocal boundary conditions (4) these are azimuthal and axial tangent components. The analysis of HC MFs with a curvilinear core – cladding boundary and discrete rotational symmetry of the cladding shows that tangential field components change their structure under moving along the boundary perimeter unlike the boundary condition (4) typical for waveguide microstructures with a continuous rotational symmetry of the core – cladding boundary. Axial field components are always tangent to the hollow core boundary while the different transverse field components become tangent to the boundary at different points which follows from the changes in the local normal vector $\vec{n}$. If the local normal vector has a coordinate $\vec{n} = (-\vec{\rho}; 0)$ at the core – cladding boundary point closest to the coordinate origin in the cylindrical coordinate system connected with the microstructure symmetry axis then its coordinate is $\vec{n} = (0; \vec{\varphi})$ at the angular point of the core – cladding boundary. Here, the vector $\vec{\rho}$ is a unit vector in the radial direction in a given coordinate system and $\vec{\varphi}$ is a unit vector in the azimuthal direction. Similar conclusions can be drawn for the tangent vector $\vec{t}$. It means that in this case the boundary conditions can be considered only as local. Thus, the air core mode formed in the determined region of the core – cladding boundary for one set of the local tangent field components cannot exist in the same form in another region of the core – cladding boundary under changed local boundary conditions.

The above process is very similar to the process of mode formation in square and triangular microstructures (Fig. 6, 7) when the boundary conditions for tangent electromagnetic field components are different in the centers of the polygon sides and in the angular domains of the polygons. As for the microstructure in Fig. 8, here, unlike in other microstructures, the low loss waveguide level is determined by the boundary regions with the predominant local ARROW mechanism.

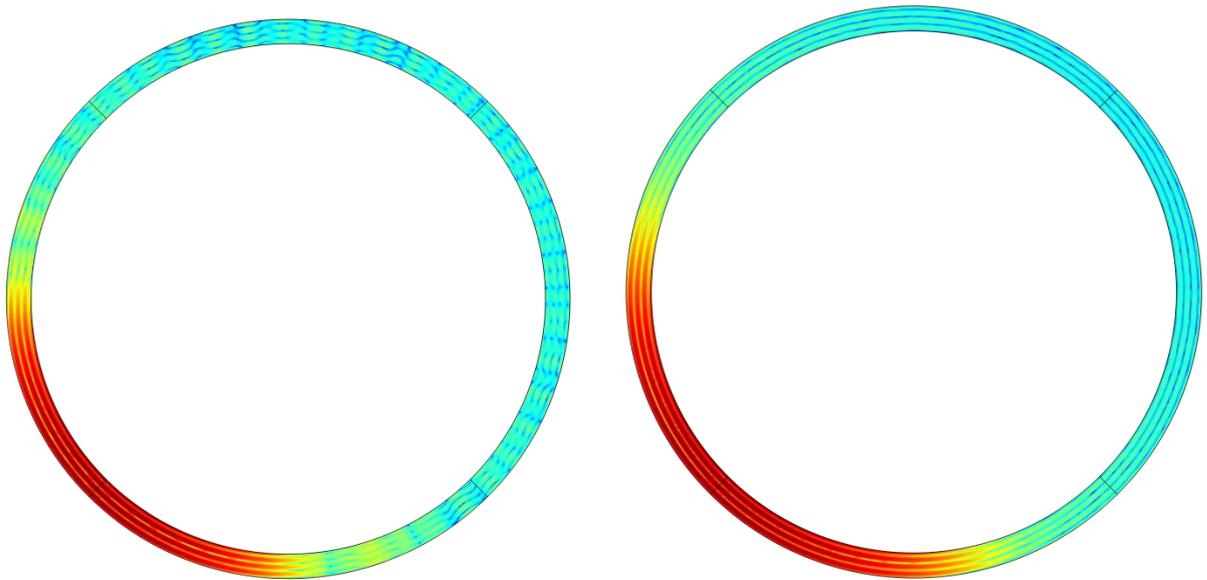

Fig. 11. The distribution of the absolute values of transverse electric field components of the air core modes in the cladding capillary wall in typical HC MF with a negative curvature core – cladding boundary: (left) the cladding capillaries touch each other; (right) the cladding capillaries don't touch and the distance between them is 7.7 μm. The part of the capillary surface closest to the air core is red.

The air core mode energy flows out in these relatively narrow boundary regions as it is in these regions that the same sets of tangent components of the air core mode fields occur. At the same time, for the microstructure in Fig. 9 the boundary conditions localization along the air core boundary is not so clear. As it was pointed out above, here the predominate process is the nonlocal application of boundary conditions in the azimuthal photonic crystal regime with the discrete rotational symmetry of the core – cladding boundary as a whole.

Note that the core – cladding boundary should be continuous or quasi continuous to observe the field distributions in the microstructure walls determined by different mechanisms of light localization. We consider the core – cladding boundary to be quasi continuous when a small clearance between the core – cladding boundary elements (capillaries) can exist and allow the air core mode fields localized in the element walls to sufficiently interact. In this case, we can introduce local normal vectors to the inner surface of the core – cladding boundary and corresponding local boundary conditions.

Direct numerical calculations show that similar distributions of the air core mode fields in the core – cladding boundary walls also occur for real HC MFs which we used in [13]. The HC MF cladding consisted of eight capillaries; the distributions of the absolute value of the transverse electric field components of the fundamental air core mode in the capillary wall are shown in Fig. 11. If the capillaries touch each other (Fig. 11(left)) the fields distribution in the capillary wall is the same as the microstructure in Fig. 8. Here, different regions of the quasi continuous boundary correspond to different boundary conditions. If the cladding capillaries are moved apart at some distance from each other then the field distribution in the capillary wall evolves into a standard ARROW distribution (Fig. 11(right)). As it was shown in [17] there is an optimal distance between the cladding capillaries where the minimal waveguide loss is observed. This phenomenon can be related to the total suppression of the azimuthal photonic crystal mechanism (contribution). Each boundary element starts acting as an individual scatterer with its own boundary conditions and its own system of the local normals (outward normals, in this case) with a further increase in the distance between the capillaries. Here, the optical characteristics of the waveguide system are determined by individual optical properties of a cladding capillary and its resonant properties are well described by the ARROW model [6] but not by the resonant properties of the core – cladding boundary as a whole.

## 4. Conclusions

It has been shown that the light localization mechanism in polygonal waveguide microstructures is different from the standard ARROW mechanism. The air core modes are determined not only by the azimuthal number (or, the azimuthal momentum of the harmonic) as in dielectric tubes but by a superposition of harmonics with different values of azimuthal momentums. A complicated fields distribution is formed in the microstructure wall and an azimuthal number degeneracy of the field components is manifested. The degeneracy of the air core mode fields occurs due to a discrete rotational symmetry of the core – cladding boundary. The interference of the harmonics forming the air core modes results in a more complicated interaction of the air core mode with the boundary in azimuthal direction. The periodic azimuthal distribution of the air core mode fields along the air core boundary leads to a decrease in the polygonal microstructures waveguide loss compared to that in circular dielectric tubes. As in the case of circular dielectric tubes, the air core mode formation mechanism is not local. The locality of the air core mode formation mechanism is manifested in polygonal microstructures with a number of sides $N < 5$. The antiresonant light reflection (which is an analog of the ARROW mechanism) occurs in the central regions of the core – cladding boundary while in the angular domains of the boundary this mechanism cannot be formed due to changing boundary conditions. Correspondingly, there occurs a significant decrease in length of the boundary parts along which the air core modes effectively radiate into the outer region which leads to the waveguide loss decrease. A similar phenomenon can occur in the curvilinear core – cladding boundary. Here, the decrease in the effective length of the air core modes and the air core boundary interaction occurs due to a 'negative curvature' of individual boundary elements. It leads to the boundary conditions localization for the tangent field components. If the curvature of the microstructure core – cladding boundary is large enough the boundary conditions localization leads to a sharp fall in the waveguide loss. We can then regulate the microstructure waveguide loss level by changing the geometry parameters of an individual boundary element.

To summarize, the standard ARROW mechanism [6] describes interactions between radiation and the periodic cladding structure which are local in the transverse direction. For the considered polygonal microstructures this locality also occurs in the azimuthal direction. In this case, the core – cladding boundary ceases to behave as a unity with equal boundary conditions along the boundary perimeter. As a result, there occur local parts of the boundary where the air core mode fields are formed due to a mechanism analogous to the ARROW mechanism.